# Resolution enhancement by extrapolation of coherent diffraction images: A quantitative study on the limits and a numerical study of non-binary and phase objects

Tatiana Latychevskaia[1]*, Yuriy Chushkin[2] and Hans-Werner Fink[1]

[1]Department of Physics, University of Zurich, Winterthurerstrasse 190, CH- 8057 Zürich, Switzerland

[2]European Synchrotron Radiation Facility, 71 Avenue des Martyrs, 38000 Grenoble, France

*Corresponding author: tatiana@physik.uzh.ch

# CONTENTS

# ABSTRACT

In coherent diffractive imaging (CDI) the resolution of the reconstructed object is limited by the numerical aperture of the experimental setup. We present here a theoretical and numerical study for achieving super-resolution by post-extrapolation of coherent diffraction images, such as diffraction patterns or holograms. We demonstrate that a diffraction pattern can unambiguously be extrapolated from only a fraction of the entire pattern and that the ratio of the extrapolated signal to the originally available signal is linearly proportional to the oversampling ratio. While there could be in principle other methods to achieve extrapolation, we devote our discussion to employing iterative phase retrieval methods and demonstrate their limits. We present two numerical studies; namely the extrapolation of diffraction patterns of non-binary and that of phase objects together with a discussion of the optimal extrapolation procedure.

# INTRODUCTION

It is an inherent property of a wave that it propagates in space and thus is spread out over some distance. In coherent diffractive imaging (Miao *et al.*, 1999), a wave scattered by an object is detected in the far-field. However, physical detectors always posses a certain size and thus capture only a limited section of the wave distribution. This instrumental restriction is analogous to the Abbe criterion, which states that the resolution is limited by the numerical aperture of the optical system given a fixed wavelength.

When the object distribution is known exactly, the far-field distribution of the wave scattered by the object is also known, and vice versa. This idea has previously been discussed by Harris in 1964 (Harris, 1964). In Harris' terminology, the far-field distribution of the wave scattered by an object is the spectrum of the object. Assuming that only part of the spectrum is captured by a detector, Harris studied the ambiguity of the spectrum beyond the cut-off frequency of the optical system, where the latter is determined by the detector size. He proved that when an object posses a finite size, its spectrum beyond the cut-off frequency is unambiguously defined. The proof is based on two theorems. The first states that the spectrum of a size-limited object is always analytical (Whittaker & Watson, 1935). The second concerns the analytical functions and states that "a function of a complex variable is determined throughout the entire Z-plane from a knowledge of its properties within an arbitrarily small region of analyticity" (Guellemin, 1951). This means that knowledge of a piece of spectrum is sufficient to determine the entire spectrum. What immediately follows from the theorem is that the entire spectrum is unambiguously defined from its section. The corollary to the analyticity theorem states that "any two functions of a complex variable whose values coincide over an arbitrarily small region of analyticity must have identical values throughout their common region of analyticity and hence must be identical. This statement is known as the identity theorem or uniqueness theorem for analytic functions." In his conclusions, Harris states that "for objects of finite angular dimensions, knowledge of the spatial frequency spectrum within the bandpass of any imaging system implies knowledge of the spatial frequency spectrum over the entire frequency domain, and hence implies *complete* knowledge of the object". Another important conclusion of Harris is that the optical resolution is therefore only limited by the noise of the system and not by some absolute criterion. With his study, Harris provided the theoretical foundation for the problem of spectrum extrapolation beyond the detector area. However, Harris considered only the case of an analytical signal, whereas a practical application of such a spectrum extension would be in the signal analysis applied to digitally sampled signals. Later, Gerchberg (Gerchberg, 1974) and Papoulis (Papoulis, 1975), studied the practical possibility of extrapolation of the spectrum of a digital signal. However, their works addressed the situation when a section of spectrum is known exactly, meaning that the amplitude and phase distributions must be known.

CDI is a relatively modern imaging technique where the structure of a non-crystalline object is recovered from its diffraction pattern (Miao et al., 1999). Since the resolution of the recovered object is defined by the highest frequencies detected in the diffraction pattern, it can be increased by employing a larger detector. However, as discussed above, in principle, the entire spectrum can be determined only from its section.

The crucial differences between the extrapolation discussed by Harris and an extrapolation of an experimental diffraction pattern are the following:

(1) As previously outlined by Harris, the problem of extrapolation has a unique solution for any analytical signal. In CDI, both signals, in the Fourier and in the object domains, are digitally sampled.

(2) A section of the spectrum must be known precisely, including its amplitude and phase distributions. In CDI, the detected section of the spectrum is not exactly known, since only amplitudes respectively intensities are measured but the phase information is lost. The missing phase can however be recovered by employing phase retrieval routines.

(3) The solution found by conventional phase retrieval methods in CDI is only an approximate solution.

It is therefore questionable, indeed a challenge to explore, whether a far-field distribution of the wave can be extrapolated only from a section of its approximate and digitally sampled distribution. Below we address this issue in more detail and study the uniqueness of the solution. We also discuss the algorithm for spectrum extrapolation and apply it to typical types of objects, providing the optimal parameters of the extrapolation routine.

The extrapolation of the signal from part of an experimental record has already been successfully demonstrated for light optical holograms and diffraction patterns (Latychevskaia & Fink, 2013b, Latychevskaia & Fink, 2013a), terahertz in-line holograms (Rong *et al.*, 2014, Rong *et al.*, 2015), simulated diffraction patterns of crystalline samples (Latychevskaia & Fink, 2015) and experimental X-ray diffraction patterns (Latychevskaia *et al.*, 2015). The achieved enhancement in resolution is at least twice that obtained from non-extrapolated patterns (Latychevskaia & Fink, 2013a, Rong et al., 2015). Such a resolution enhancement beyond the Abbe limit has quantitatively been evaluated (Latychevskaia & Fink, 2013a). The extrapolation of simulated experimental diffraction pattern of binary objects has also been studied (Latychevskaia & Fink, 2013a). The extrapolation of an experimental X-ray diffraction pattern of a non-binary (mainly binary with some faint structure) object has recently been achieved (Latychevskaia et al., 2015). While the extrapolation of holograms of phase shifting objects has been successfully demonstrated in terahertz holography (Rong et al., 2014, Rong et al., 2015), extrapolations of diffraction patterns of absorbing and phase shifting objects have not been studied so far, and are part of this work.

## UNIQUENESS OF THE SOLUTION

### Uniqueness of the solution in conventional phase retrieval

Previous theoretical studies have demonstrated that the phase problem can be uniquely solved in two dimensions (Bruck & Sodin, 1979, Bates, 1982). In general, a unique solution to a set of equations is achieved when the number of unknowns does not exceed the number of equations. This notion was considered by Miao et al when studying the uniqueness of an object reconstruction in one, two and three dimensions obtained from the object's diffraction pattern by phase retrieval (Miao *et al.*, 1998). In this section, we address this notion in further detail. Given the density of an object $f(\vec{r})$, its Fourier transform $F(\vec{k})$ is given by

$$F(\vec{k}) = \int_{-\infty}^{+\infty} f(\vec{r}) \exp\left(-i\vec{k}\cdot\vec{r}\right) d\vec{r} \quad , \tag{1}$$

where $\vec{r}$ represents the spatial coordinates in the object domain, and $\vec{k}$ is the coordinate in Fourier domain. In a numerical analysis, the object and its Fourier transform are replaced by arrays. By using conventional sampling, we rewrite Eq. (1) as

$$F(\vec{k}\,') = \sum_{\vec{r}\,'=0}^{N-1} f(\vec{r}\,') \exp\left(-2\pi i\vec{k}\,'\cdot\vec{r}\,'/N\right) \quad , \tag{2}$$

where $\vec{r}\,'$ and $\vec{k}\,'$ are digitised in Eq. (2) and stand for pixels that range in each dimension from 0 to $N-1$, where $N$ is the number of pixels of the detector in one dimension. Since only the magnitude of the Fourier transform is measured, the acquired data correspond to

$$\left|F(\vec{k}\,')\right| = \left|\sum_{\vec{r}\,'=0}^{N-1} f(\vec{r}) \exp\left(-2\pi i\vec{k}\,'\cdot\vec{r}\,'/N\right)\right|, \tag{3}$$

respectively

$$\left| \sum_{\vec{r}'=0}^{N-1} f(\vec{r}') \exp\left(-2\pi i \vec{k}' \cdot \vec{r}' / N\right) \right| - \left| F(\vec{k}') \right| = 0. \tag{4}$$

Equation (4) is in fact a set of equations, and the problem is to find the phase for $f(\vec{r}')$ at each pixel, whereby a pixel represents an element of the $f(\vec{r}')$ and $F(\vec{k}')$ arrays.

We consider a $\gamma$-dimensional case, where $\gamma = 1, 2, 3$, etc. We assume that $f(\vec{r}')$ is a complex-valued distribution. For $\gamma$-dimensional complex-valued objects, the total number of equations is $N^{\gamma}$. The total number of the pixels with unknown values in the object domain is $N^{\gamma}$, and the total number of unknown variables is $2N^{\gamma}$, since each pixel has two unknown variables: the real part and the imaginary part.

To decrease the number of unknown variables, an object support consisting of pre-known values is applied. To determine how many known-support-valued pixels of $f(\vec{r}')$ are necessary for solving Eq. (4), we introduce the oversampling ratio $\Omega$ defined as (Miao et al., 1998):

$$\Omega_0 = \frac{\text{total pixel number}}{\text{number of pixels with unknown-value}} = \frac{N}{N_{\text{unknown}}}, \tag{5}$$

This gives the number of pixels with unknown values:

$$N_{\text{unknown}}^{\gamma} = \left(\frac{N}{\Omega}\right)^{\gamma}, \tag{6}$$

and the number of unknowns in the object domain:

$$2N_{\text{unknown}}^{\gamma} = 2\left(\frac{N}{\Omega}\right)^{\gamma}. \tag{7}$$

where $\Omega$ is the linear oversampling ratio. The set of equations in (4) can be solved when the number of unknowns is equal to or less than the number of equations

$$2N_{\text{unknown}}^{\gamma} = 2\left(\frac{N}{\Omega}\right)^{\gamma} \leq N^{\gamma}, \tag{8}$$

which requires the oversampling ratio to be

$$\Omega \geq 2^{1/\gamma}. \tag{9}$$

## Uniqueness of the extrapolation of amplitudes in the diffraction pattern

In this example, we consider a diffraction pattern with only the amplitudes in the Fourier domain being known. The diffraction pattern is sampled with $N^{\gamma}$ pixels and is subject to an extrapolation to $(N_{\text{ex}})^{\gamma}$ pixels. We introduce the extrapolation factor $\varepsilon$ as

$$\varepsilon = \frac{N_{\text{ex}}}{N}. \tag{10}$$

The set of equations in (4) can be written for $(N_{ex})^\gamma$ pixels, thus giving $N_{ex}^\gamma = (\varepsilon N)^\gamma$ equations. The number of unknowns is given by the sum of unknowns in the Fourier domain and in the object domain. The number of *pixels* with unknown values in the object domain is given by $\left(\dfrac{N_{ex}}{\Omega}\right)^\gamma$, and the number of *unknowns* in the object domain is $2\left(\dfrac{N_{ex}}{\Omega}\right)^\gamma$ because each pixel has two unknown variables: the real and the imaginary part. In addition to these unknowns, the pixels added in the Fourier domain during the extrapolation, are also counted as unknowns. Since only their absolute values contribute to Eq. (4), the additional number of unknowns is the number of pixels added during extrapolation $N_{ex}^\gamma - N^\gamma$. In order to solve an extrapolated diffraction pattern, the number of unknowns must not exceed the number of equations:

$$2\left(\frac{N_{ex}}{\Omega}\right)^\gamma + \left(N_{ex}^\gamma - N^\gamma\right) \le N_{ex}^\gamma, \tag{11}$$

which shortens to

$$\varepsilon \le \frac{\Omega}{2^{1/\gamma}}. \tag{12}$$

When the inequality in Eq. (12) is fulfilled, a unique solution of the object distribution as well as uniquely defined values at extrapolated pixels can be found. The relation between a typical oversampling ratio $\Omega$ and the related extrapolation ratio $\varepsilon$ for $\gamma = 2$ is shown in Fig. 1.

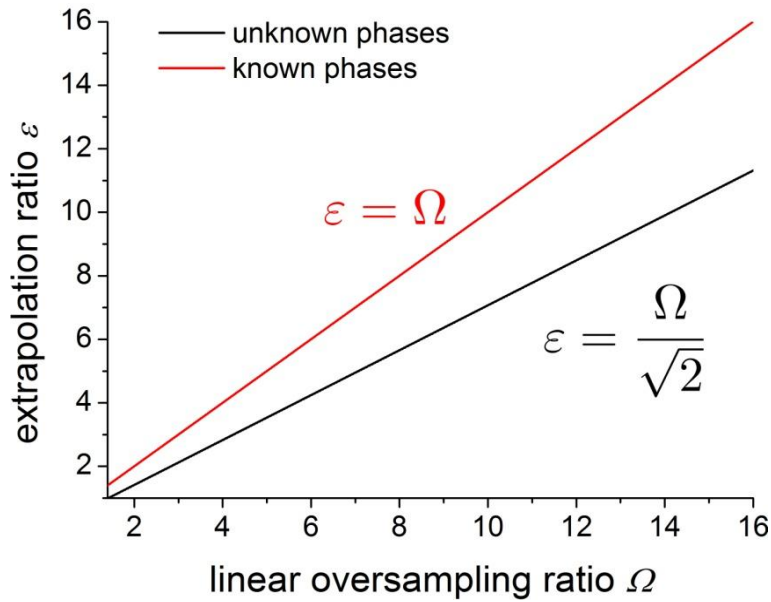

Fig. 1. Extrapolation ratio $\varepsilon$ as function of the oversampling ratio $\Omega$ for a two-dimensional diffraction pattern ($\gamma = 2$). Black line shows the function $\varepsilon(\Omega)$ when only the amplitudes of the complex-valued wavefront in the far-field (diffraction pattern) are known but the phases are not known. Red line shows the function $\varepsilon(\Omega)$ when both, amplitudes and phases of the complex-valued wavefront in the far-field are known.

## Uniqueness of the extrapolation of the complex-valued wavefront in the far-field

In this example, we assume that a diffraction pattern has been subject to phase retrieval, and that for $N^\gamma$ pixels, the amplitudes and the phases are known. This is analogous to the requirement of the Gerchberg-Papoulis algorithm, that the spectrum must be known (Gerchberg, 1974, Papoulis, 1975). Then, the set of equations can be re-written for complex-valued functions

$$\sum_{\vec{r}'=0}^{N-1} f(\vec{r})\exp\left(-2\pi i \vec{k}'\cdot \vec{r}'/N\right) - F(\vec{k}') = 0 \qquad (13)$$

and the set of equations for an extrapolated diffraction pattern is:

$$\sum_{\vec{r}'=0}^{N_{ex}-1} f(\vec{r}')\exp\left(-2\pi i \vec{k}'\cdot \vec{r}'/N\right) - F(\vec{k}') = 0. \qquad (14)$$

The number of equations in (14) is $2N_{ex}^\gamma = 2\left(\varepsilon N\right)^\gamma$, where the factor 2 accounts for the real and imaginary parts. The number of unknowns is given by the sum of unknowns in the Fourier domain and in the object domain. The number of the *pixels* with unknown values in the object domain is given by $\left(\dfrac{N_{ex}}{\Omega}\right)^\gamma$, and the number of *unknowns* in the object domain is $2\left(\dfrac{N_{ex}}{\Omega}\right)^\gamma$. The number of *pixels* with unknown values in the Fourier domain is given by $\left(N_{ex}^\gamma - N^\gamma\right)$, which gives the number of *unknowns* in the Fourier domain $2\left(N_{ex}^\gamma - N^\gamma\right)$. Thus, for a uniquely defined extrapolation of the complex-valued wavefront in the far-field, the number of unknowns must not exceed the number of equations:

$$2\left(\frac{N_{ex}}{\Omega}\right)^\gamma + 2\left(N_{ex}^\gamma - N^\gamma\right) \le 2N_{ex}^\gamma, \qquad (15)$$

which simplifies to:

$$\varepsilon \le \Omega. \qquad (16)$$

In Figure 1, the range of the extrapolation ratio $\varepsilon$ as a function of $\Omega$ is shown for a two-dimensional diffraction pattern, $\gamma = 2$. Here, higher values of the extrapolation ratio $\varepsilon$ can be achieved when comparing it to the case of unknown phases in the Fourier domain.

This consideration suggests that a diffraction pattern can unambiguously be extrapolated from its section. While this is an important insight by itself, another important question is how to find this solution. In principle, there can be various ways of solving a set of equations. For example, if a diffraction pattern consists of just a few pixels, the set of equations could be solved even by hand. Here we want to emphasise that there must exist many ways of solving the set of equations, and we are not necessarily bound to iterative methods. Perhaps there will be more effective ways suggested in the future. The proof that a diffraction pattern can unambiguously be extrapolated from its section does not depend on the specific way of obtaining such a solution. Since current methods of phasing diffraction patterns are limited to the iterative phase retrieval (Fienup, 1982), we shall also employ these routines for extrapolating diffraction patterns.

Obviously, an extrapolation cannot be extended to infinity; another fundamental limit, namely the wavelength of the radiation employed, ultimately limits the achievable resolution.

## NON-BINARY REAL-VALUED OBJECT

## Simulation of diffraction pattern

In a simulation, the diffraction pattern is computed as the squared amplitude of the Fourier transform of the object distribution. In reality, a wave scattered by a physical object exists beyond the detector area. In simulations, the Fourier transform of the object distribution in the far-field exhibits a strictly defined field of view: $N\Delta_k = \dfrac{2\pi}{\Delta_O}$, where $N$ is the number of pixels, $\Delta_k$ is the pixel size in the Fourier domain, $k = \dfrac{2\pi}{\lambda}$, $\lambda$ is the wavelength and $\Delta_O$ is the pixel size in the object domain.

The spectrum of a size-limited object is infinite. Thus, the far-field distribution of a wavefront scattered by a size-limited object also spreads out to infinity. The numerical fast Fourier transform of a digitised object distribution also gives rise to the whole spectrum, but it is represented within the simulation window, which is restricted by the highest frequency $f_0$. The spectrum of a size-limited object is infinite and contains frequencies that are higher than $f_0$. For example, a sharp edge of an object can be well represented as an infinite sum of cosine functions. However, finite sampling of the object leads to the situation where some of these cosines are undersampled, or "aliased". The correct frequency related to such cosine components lies outside the spectrum of the computational window limited by $f_0$, but due to the undersampling this cosine component will re-appear at lower frequency, which is within the computational window. In this way, spectral frequency components higher than $f_0$ will be "wrapped" around the edges of the computational window and thus reappear in the spectrum computational window. To avoid this wrapping effect, it is better to select a test object without sharp edges, which does not have too many high-order frequencies in its spectrum. Also, typically only the central part of the simulated spectrum is considered. The "wrapping" effect has previously been discussed in the literature, for example by Guizar-Sicairos and Fienup (Guizar-Sicairos & Fienup, 2008). To minimise the presence of the wrapping effect in the simulated diffraction pattern, usually only some central part of the simulated image is selected. In this work, only the central $N/2 \times N/2$ pixel region of an $N \times N$ simulated diffraction pattern is considered.

The test object is the well-known image of Lena frequently used in digital image analysis. The original image, sampled with $256 \times 256$ pixel, is shown in Fig. 2(a). The object was formed (1) by setting the amplitudes of the pixels to range from 0 to 1, which corresponds to the amplitude of a realistic transmission function (Latychevskaia & Fink, 2007), and (2) by zero-padding the Lena image by up to $2048 \times 2048$ pixel, thus providing an oversampling ratio $\Omega = 8$. Then, the far-field distribution is created by calculating the Fourier transform of the object distribution, and thus the obtained complex-valued distribution is sampled with $2048 \times 2048$ pixel. The central $1024 \times 1024$ pixels area of the simulated complex-valued distribution is selected to produce a correctly simulated diffraction pattern. The reconstruction obtained by an inverse Fourier transform of this complex-valued distribution exhibits some resolution loss; see Fig. 2(b). And even more severe loss of resolution occurs when only the central $512 \times 512$ pixel region is selected; see the reconstruction shown in Fig. 2(c). Cropping of the far-field distribution to $N_{crop} \times N_{crop}$ pixel does not change the oversampling ratio, but reduces the resolution by a factor of $N/N_{crop}$. The object distributions, obtained by inverse Fourier transforms of the complex-valued far-field distributions shown in Fig. 2(b) and (c) are thus the "ideal" reconstructions that can be expected from the cropped far-field distribution when both the amplitude and phase distribution in the far-field are known and only the size of the far-field distribution is limited.

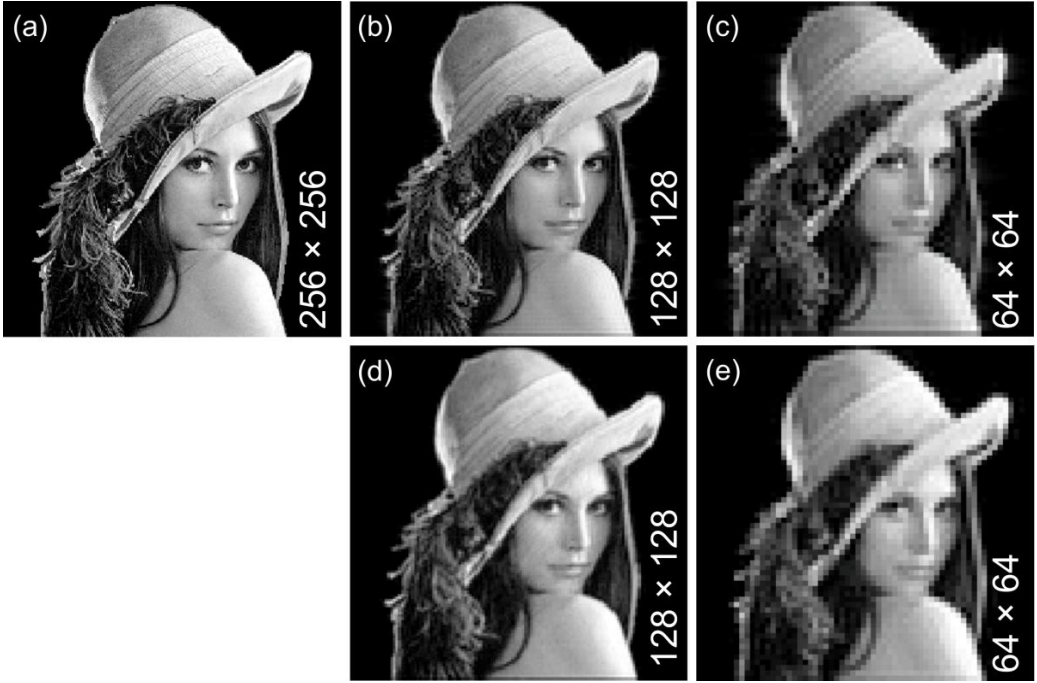

Fig. 2. Reconstructions obtained a from diffraction pattern of a non-binary object. (a) Original object distribution, Lena image, sampled with $256 \times 256$ pixels. (b) Amplitude of the reconstruction obtained from a $1024 \times 1024$ pixels complex-valued far-field distribution. (c) Amplitude of the reconstruction obtained from a $512 \times 512$ pixels complex-valued far-field distribution. (d) Average of 50 reconstructions obtained by conventional phase retrieval from a $1024 \times 1024$ pixels diffraction pattern. (e) Average of 50 reconstructions obtained by conventional phase retrieval from a $512 \times 512$ pixels diffraction pattern.

## Conventional phase retrieval procedure

Conventional phase retrieval employs an iterative approach. A typical algorithm is sketched in Fig. 3. The diffraction pattern is reconstructed by the shrinkwrap algorithm (Marchesini *et al.*, 2003) based on the hybrid input–output (HIO) algorithm (Fienup, 1982) with the feedback parameter $\beta = 0.9$. In the first iteration, random phases are combined with the measured amplitudes. The real part of the inverse Fourier transform of the resulting complex-valued distribution provides the object distribution $g_k(x, y)$ ($k = 1$ for the first iteration). Then, the iterative procedure is applied, which includes the following steps:

(i) A Fourier transform of $g_k(x, y)$ provides the complex-valued distribution $G_k(X, Y)$, where $(X, Y)$ are the coordinates in the detector plane.

(ii) The iterated amplitudes $\left| G_k(X, Y) \right|$ are replaced with the measured amplitudes $G_{\exp}(X, Y)$ and thus an updated distribution $G_k^{'}(X, Y)$ is formed:

$$G_k^{'}(X, Y) = G_{\exp}(X, Y) \exp\left( \mathrm{iArg}\left( G_k(X, Y) \right) \right)$$

(iii) The real part of the inverse Fourier transform of $G_k^{'}(X, Y)$ provides $g_k^{'}(x, y)$.

(iv) The object distribution $g_{k+1}(x, y)$ is calculated as follows:

$$g_{k+1}(x, y) = \begin{cases} g_k^{'}(x, y) & \text{if } (x, y) \in S_1, \\ g_k(x, y) - \beta g_k^{'}(x, y) & \text{if } (x, y) \notin S_1, \end{cases} \tag{17}$$

where $S_1$ denotes the set of the points at which the coordinates $(x, y)$ are within the object support region and $\mathrm{Re}\left\{ g_k^{'}(x, y) \right\} > 0$. The resultant $g_{k+1}(x, y)$ is plugged into the next iteration starting at (i).

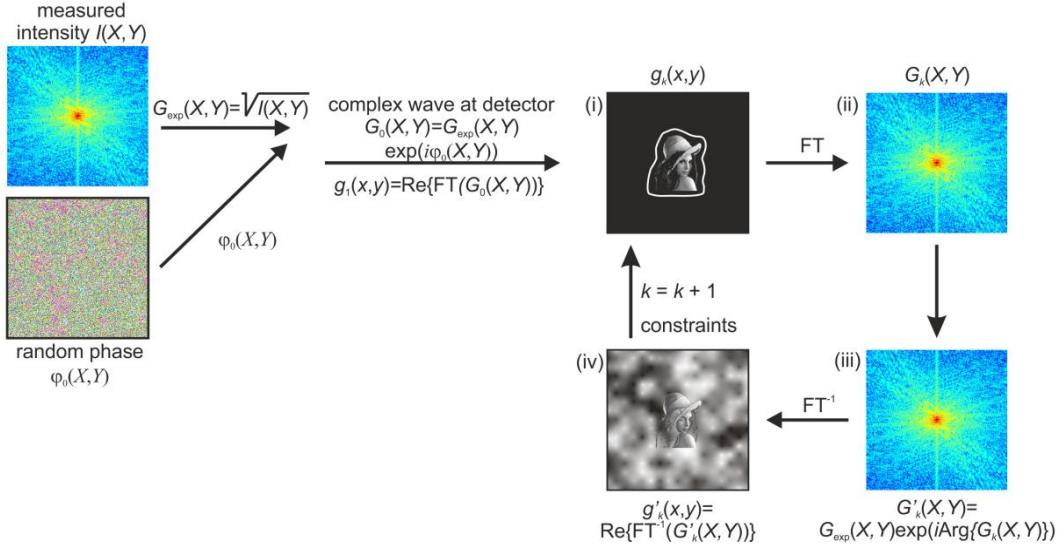

Fig. 3. Illustration of the conventional iterative phase retrieval scheme for a real-valued object.

The object support $S_1$ is calculated as described in the shrinkwrap algorithm [11]. The Fourier transform of the experimental diffraction pattern provides the autocorrelation of the object and with a threshold of $tr_1$ at its maximum, it delivers the object support $S_1$ for the first iteration. After every 20th iteration, the object support $S_1$ is updated by calculating the convolution $g_k(x, y)$ with a two-dimensional Gaussian distribution $\exp\left(-\left(x^2 + y^2\right)/2\sigma^2\right)$ and a threshold of the amplitude of the result at $tr_2$ of its maximum. The standard deviation of the Gaussian distribution $\sigma$ was gradually reduced from $\sigma_1$ to $\sigma_2$.

The quality of the retrieved complex-valued amplitudes in the detector plane is evaluated by calculating the misfit between the measured and the iterated amplitudes, or the error:

$$E = \frac{\sum_{i,j=1}^{N} \left| G_{\exp}(i, j) - \left| G_{\mathrm{it}}(i, j) \right| \right|}{\sum_{i,j=1}^{N} \left| G_{\exp}(i, j) \right|},$$ (18)

where $G_{\exp}(i, j)$ are the experimentally measured amplitudes at the detector, $\left| G_{\mathrm{it}}(i, j) \right|$ are the iterated amplitudes, $i$ and $j$ are the pixel numbers $i, j = 1...N$.

Next, fifty iterated distributions $G_{\mathrm{it}}(i, j)$ with the least error $E$ are selected, and each one is stabilised with an additional 1000 iterations with the error-reduction (ER) algorithm (Fienup, 1982). Two constraints are applied in the object domain. The first is the support constraint:

$$g_{k+1}(x, y) = \begin{cases} g_k(x, y) & \text{if } (x, y) \in S_2, \\ 0 & \text{if } (x, y) \notin S_2, \end{cases}$$ (19)

where $S_2$ is the objects support. The object support $S_2$, is calculated as follows. The amplitude of the reconstructed object distribution is convolved with a two-dimensional Gaussian distribution with the standard deviation $\sigma_3$. The support $S_2$ is obtained by calculating a threshold of the amplitude of the result of the convolution at $tr_3$ of its maximum. The support $S_2$ is updated at each iteration. The second constraint is the threshold constraint:

$$g_{k+1}(x, y) = \begin{cases} g_{k+1}(x, y) & \text{if } g_{k+1}(x, y) < T \\ T & \text{if } g_{k+1}(x, y) \geq T \end{cases}, \qquad (20)$$

which is based on the notion that the amplitude of the transmission function cannot exceed some certain threshold value $T$ (Latychevskaia & Fink, 2007). The threshold value $T$ is related to the maximum of the transmission function of the sample; in our simulated sample it was set to $T = 1$.

## Conventional phase retrieval of non-binary object

The diffraction pattern of size $1024 \times 1024$ pixel was retrieved with the following parameters in the HIO algorithm: $tr_1 = 3\%$, $tr_2 = 5\%$, $\sigma_1 = 2.5$ and $\sigma_2 = 1.7$. 1000 iterative runs were carried out. Each iterative run included 2000 iterations. Next, 1000 iterations of the ER algorithms are performed for 50 reconstructions with the least error $E$ and a meaningful reconstruction. In the ER algorithm, only the amplitude distribution is considered in the object domain, the phase distribution is set to zero. The parameters of the ER algorithm are: $tr_3 = 2\%$, $\sigma_3 = 1.7$. Since the reconstructed distribution is sampled with $1024 \times 1024$ pixel and not with $2048 \times 2048$ pixel as per the original distribution, according to the Parseval's theorem the threshold was set to $T = 4$. The result of averaged 50 reconstructions is shown in Fig. 2(d).

The diffraction pattern of the Lena image sampled with $512 \times 512$ pixel was retrieved with the following parameters in the HIO algorithm: $tr_1 = 3\%$, $tr_2 = 5\%$, $\sigma_1 = 1.7$ and $\sigma_2 = 1.2$. 1000 HIO iterative runs were carried out, each with 2000 iterations. 50 reconstructions with the least error $E$ were further iterated with the ER algorithm. In the ER algorithm, only the amplitude distribution was considered in the object domain, and the phase distribution is set to zero. The parameters of ER algorithm were: $tr_3 = 2\%$, $\sigma_3 = 1.2$, $T = 16$. The result of the averaged 50 reconstructions is shown in Fig. 2(e).

It is worth noting that the reconstructions obtained by the inverse Fourier transform of the complex-valued far field distributions, shown in Fig. 2(b) and (c), are better resolved than the reconstructions obtained by phase retrieval shown in Fig. 2(d) and (e).

# Extrapolation of a diffraction pattern of a non-binary real-valued object

The procedure of extrapolation is very similar to that described elsewhere (Latychevskaia & Fink, 2013b, Latychevskaia & Fink, 2013a). In the extrapolation, the pixel size in the detector (Fourier) domain does not change, but the number of pixels is increased. Thus, the reconstructed object area size remains the same as obtained from a conventional reconstruction, but it is sampled with an increased number of pixels.

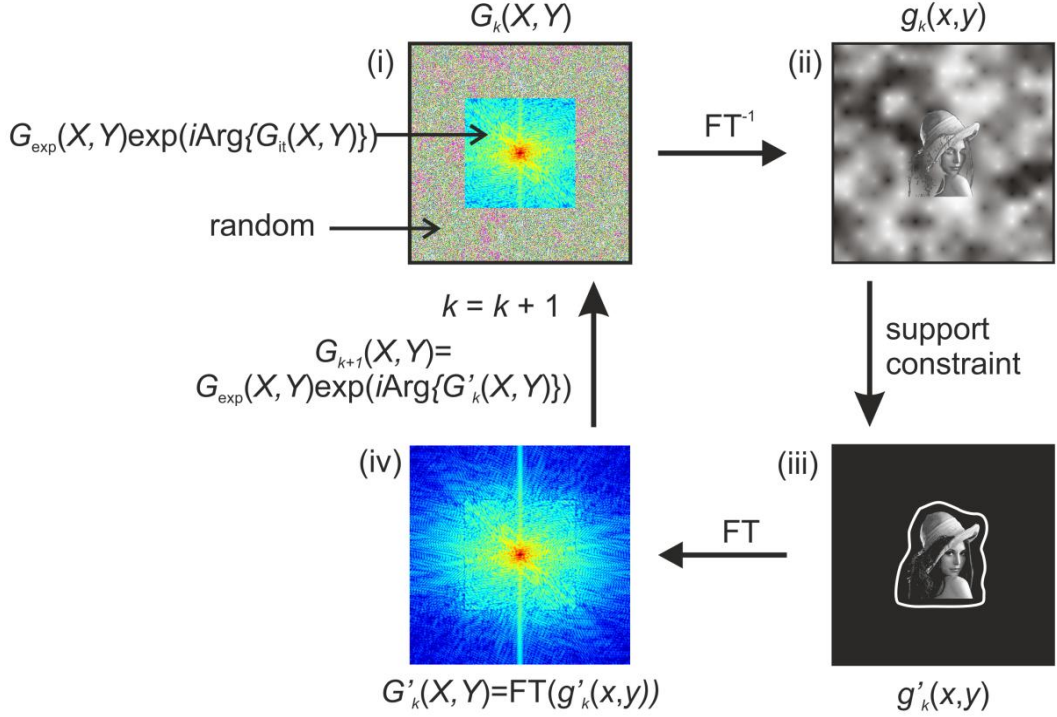

Fig. 4. Illustration of the extrapolation iterative scheme.

The extrapolation is performed by an iterative routine based on the ER algorithm (Fienup, 1982) which is sketched in Fig. 4. At the first iteration, the complex-valued wavefront distribution in the Fourier domain, previously obtained by the conventional phase retrieval routine, is padded with random complex-valued numbers up to $1024 \times 1024$ pixel. The amplitude of the padding signal is selected to be randomly distributed, ranging from 0 to 200. It has been shown that the amplitude of the initial padding only affects the speed of convergence towards the reconstruction, but not the quality of the reconstruction (Latychevskaia et al., 2015). The phase of the padding signal is randomly distributed from $-\pi$ to $+\pi$. Two constraints are applied in the object domain. The first constraint is the object support $S_{\text{Extrapolation}}$ calculated as follows. The object distribution is reconstructed by conventional phase retrieval and its amplitude are re-sampled with $1024 \times 1024$ pixel. Next, the convolution of the result with a two-dimensional Gaussian distribution is calculated; the standard deviation of the Gaussian distribution $\sigma_{\text{Extrapolation}} = 1.2$. The support $S_{\text{Extrapolation}}$ is obtained by calculating a threshold of the amplitude of the result of the convolution at 3% of its maximum. The support $S_{\text{Extrapolation}}$ is kept the same during the whole extrapolation procedure. It should be noted that an accurate and tight support should be used; otherwise some artefacts, such as black dots of 1–2 pixel in width might appear in the reconstruction. The second constraint in the object domain is the threshold $T_{\text{Extrapolation}} = 4$. In the Fourier domain, the amplitudes are replaced with the experimental values but are updated in the padding region. The phase distribution in the Fourier domain is updated following each iteration. In total, 1000 iterations are performed. Each of the 50 reconstructions is extrapolated individually by employing the same procedure. The individual reconstructions are then aligned and averaged. The extrapolated diffraction pattern is then obtained as the squared absolute value of the Fourier transform of the averaged reconstruction.

In order to check the potential of the conventional phase retrieval methods for extending the signal beyond the measured fraction, we first tested the iterative extrapolation routine using a 512 × 512 pixel simulated diffraction pattern with a known phase distribution in the far-field detector plane. Figure 5(a) shows the resulting reconstruction obtained from a 1024 × 1024 pixel extrapolated diffraction pattern. The reconstruction exhibits a higher resolution compared to the "ideal" reconstruction obtained from a 512 × 512 pixel complex-valued far-field distribution, compare Fig. 5(a) and Fig. 2(c). However, the reconstruction exhibits lower resolution compared to the "ideal" reconstruction obtained from the 1024 × 1024 pixel complex-valued far-field distribution, compare Fig. 5(a) and Fig. 2(b).

In addition, the reconstruction suffers from a superimposed Moire-like pattern. To understand the origin of this pattern, we performed additional simulations and arrived at the following conclusions. This artefact does not depend on the oversampling ratio. The reconstruction before the extrapolation can be artefact-free, but the Moire-like pattern appears already after the first iteration of the extrapolation. There is no such pattern for binary objects where a constant threshold during the extrapolation can be set correctly (Latychevskaia & Fink, 2013a, Latychevskaia et al., 2015), but this artefact can appear even for an almost binary object when the threshold is set incorrectly (Latychevskaia et al., 2015). We believe that setting the threshold more precisely, and to some variable value for non-binary objects should eliminate this artefact, and we are planning to investigate this issue further. The Moire-like pattern can be reduced when a great number of reconstructions are averaged. Fig. 5(b) shows the related extrapolated diffraction pattern. The edges of the original fraction of the diffraction pattern are apparent and the extrapolated signal does not perfectly match the original signal, as a comparison of Fig. 5(b) and Fig. 5(f) shows.

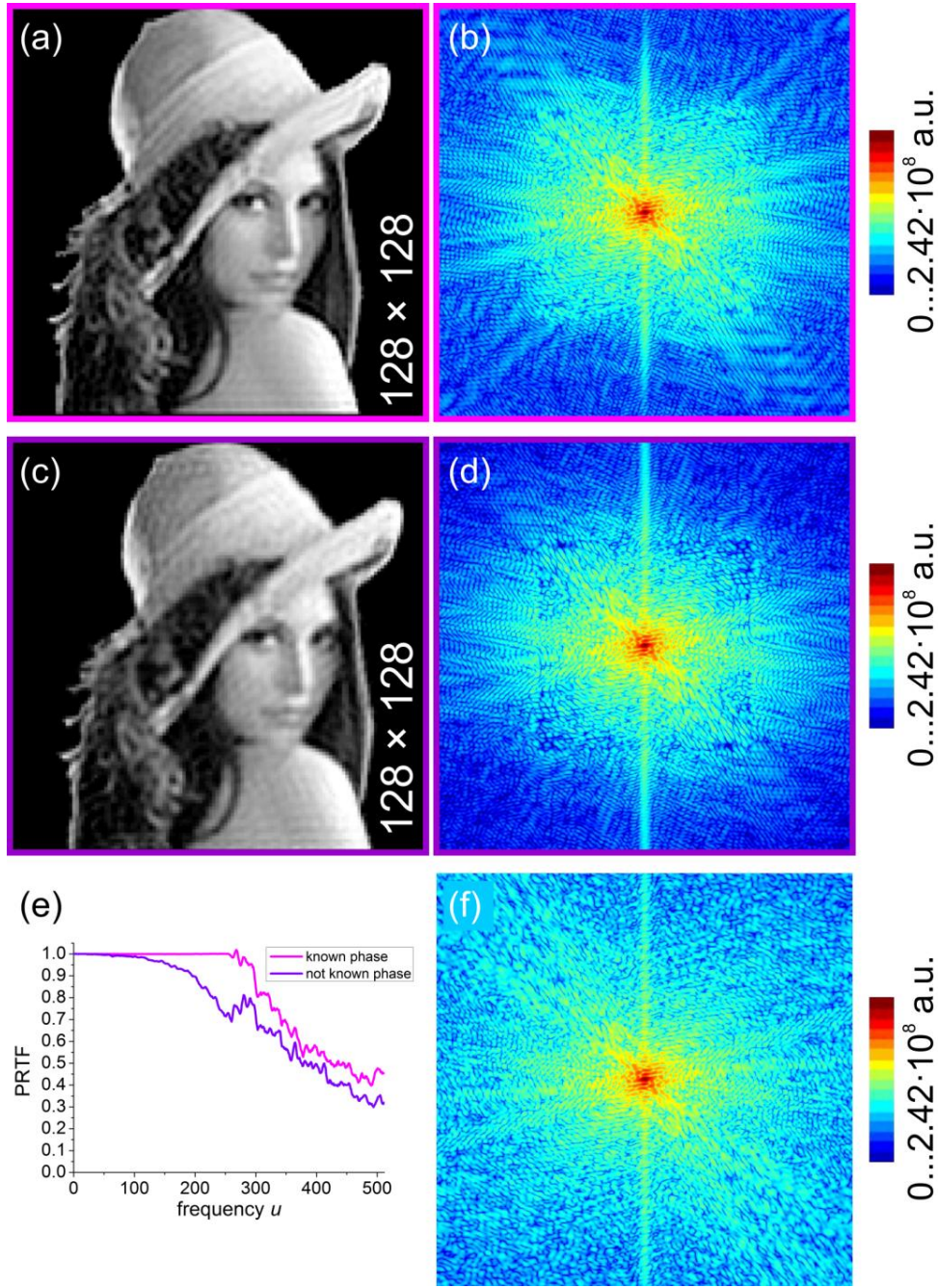

Fig. 5. Extrapolation of a non-binary object. (a) Average of 50 reconstructions obtained by extrapolation from a $512 \times 512$ pixels to a $1024 \times 1024$ pixels pattern when the phases in the far-field are known and (b) the related extrapolated diffraction pattern. (c) Average of 50 reconstructions obtained by the extrapolation from a $512 \times 512$ pixels to a $1024 \times 1024$ pixels pattern when the phases are not known and (d) the related diffraction pattern. (e) PRTF calculated for the diffraction patterns shown in (b) and (d). (f) Original diffraction pattern sampled with $1024 \times 1024$ pixels.

Next, we extrapolate a $512 \times 512$ pixel simulated diffraction pattern without known phases in the far-field detector plane. Figure 5(c) shows the resulting reconstruction obtained from a $1024 \times 1024$ pixel diffraction pattern. The reconstruction is similar to that obtained with the known phases, compare Fig. 5(a) and (c). It exhibits a higher resolution when compared to the "ideal" reconstruction obtained from a $512 \times 512$ pixel complex-valued far-field distribution, compare Fig. 5(c) and Fig. 2(c). However, it exhibits a lower resolution compared to the "ideal" reconstruction obtained from a $1024 \times 1024$ pixel complex-valued far-field distribution, compare Fig. 5(c) and Fig. 2(b). Figure 5(d) shows the related extrapolated diffraction pattern. The border between the original diffraction pattern and the extrapolated part is more clearly apparent than in the case when the phases in the far-field are known, compare Fig. 5(b) and Fig. 5(d).

The resolution is quantitatively estimated by calculating the Phase Retrieval Transfer Function (PRTF), which estimates the relation of the recovered amplitudes to the experimentally measured amplitudes (Shapiro *et al.*, 2005, Chapman *et al.*, 2006):

$$\text{PRTF}(u) = \frac{\left| \left\langle G_{\text{it}}(u) \right\rangle \right|}{\left| G_{\text{exp}}(u) \right|}, \tag{21}$$

where $u = \sqrt{\left(i - N/2\right)^2 + \left(j - N/2\right)^2}$ is the spatial frequency coordinate (in pixels), $i$ and $j$ are the pixel numbers in the detector plane, and $\left\langle ... \right\rangle$ denotes averaging over the complex-valued iterated amplitudes. The calculated PRTFs are shown in Fig. 5(e). Both PRTFs indicate a significant deviation of the recovered amplitudes from the true amplitudes in agreement with the visual quality of the reconstructions. This leads to the following conclusions: (1) even when the complex-valued numbers in the far-field are known, the extrapolation by an iterative procedure does not recover well the original values. It is interesting to note that when the phases are unknown, the extrapolated amplitudes are as close to the original amplitudes as in the case of the known phases. (2) The missing phases are less of a problem for extrapolation than the choice of the extrapolation procedure. Obviously, the iterative routines provide only an approximately extrapolated signal that is insufficient to fully recover the resolution. Seldin and Fienup showed that iterative algorithms can stagnate due to the ambiguous solutions that are very close to the given object (Seldin & Fienup, 1990).

## PHASE-SHIFTING OBJECT

## Simulation of diffraction pattern

In this section we study the extrapolation of a diffraction pattern of a phase object. The object in the form of bars was created as follows. The total object area was selected to be $1024 \times 1024$ pixel. Three bars, each of P pixel width separated by P pixels, are created. The three bar patterns are repeated five times. The following P values are selected: 6, 5, 4, 3 and 2 pixels. The amplitude of the transmission function was set to 0 outside the bars and to 1 inside the bars. The phase of the transmission function was set to 0 outside the bars and to $0.1\pi$ and $0.15\pi$ inside the bars. The amplitude and phase distributions are shown in Fig. 6(a). The total size of the object was $30 \times 180$ pixel, thus giving an oversampling ratio of about 5.7. The diffraction pattern, shown in Fig. 6(b), was simulated as an analytical solution of the diffraction at a rectangular aperture to avoid wrapping and aliasing effects described above. Then, the simulated diffraction pattern was cropped to $512 \times 512$ pixel, and its distribution is shown in Fig. 6(c). Thus, the finest features that should be reconstructed from the $512 \times 512$ pixel diffraction pattern are the set of thinnest bars that have both the width and separation between them of 1 pixel.

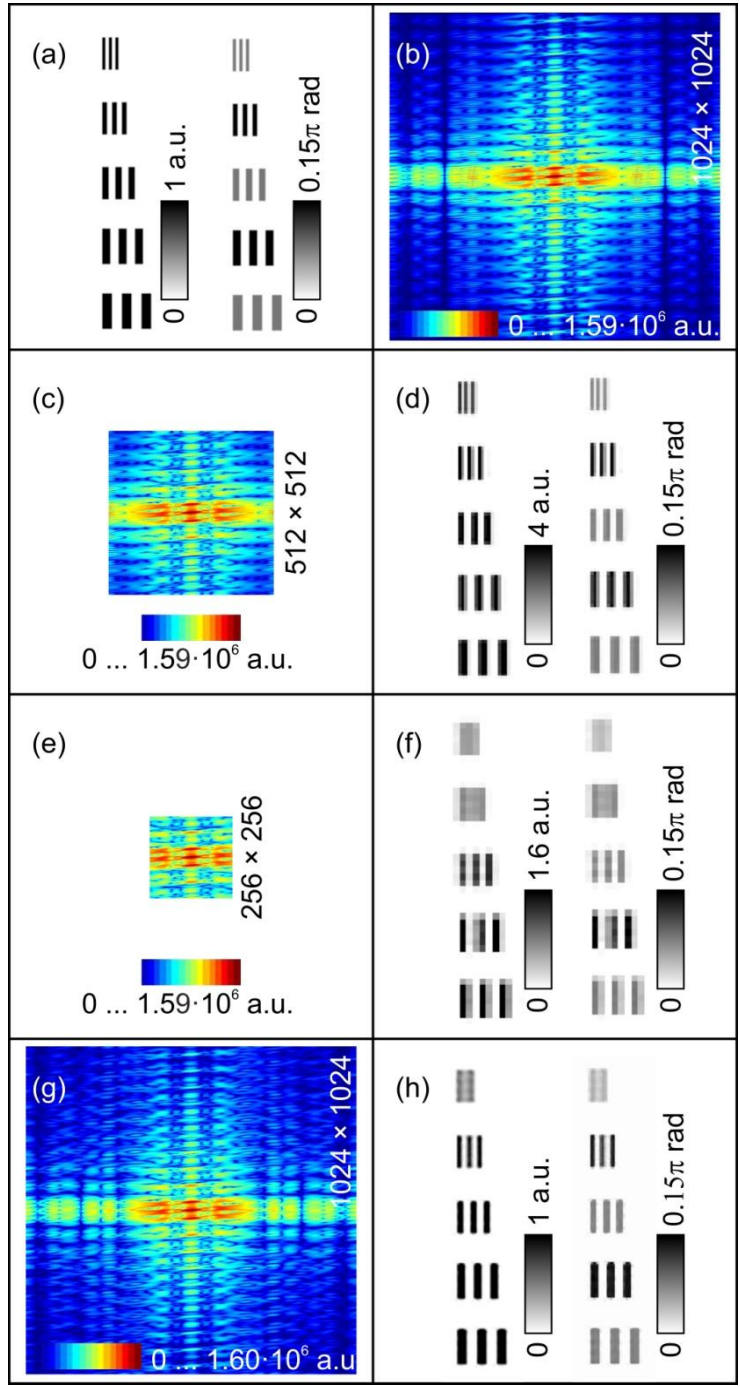

Fig. 6. Reconstructions obtained from an extrapolated diffraction pattern of a phase object. (a) Original object distribution, the transmission distribution is shown on the left and the phase distribution is shown on the right. (b) 1024 × 1024 pixels diffraction pattern simulated as analytical solution of diffraction on rectangular apertures. (c) Simulated diffraction pattern cropped to 512 × 512 pixels and (d) the related reconstruction obtained by averaging over 50 reconstructions obtained with the conventional phase retrieval. (e) Simulated diffraction pattern cropped to 256 × 256 pixels and (f) the related reconstruction obtained by averaging over 50 reconstructions obtained with the conventional phase retrieval. (g) Diffraction pattern extrapolated from 256 × 256 pixels to 1024 × 1024 pixels and (h) the related reconstruction.

# Conventional phase retrieval

The simulated diffraction pattern of $512 \times 512$ pixel in size was reconstructed by conventional phase retrieval employing 1000 iterations with the shrinkwrap algorithm, followed by 1000 iterations with the ER algorithm, as described above. An additional constraint was implemented in the object domain, which accounts for phase shifting properties of the object. In Eqs. (17) and (19), the object support $S$ included the additional restriction on the phase: $-0.2\pi < \varphi < 0.2\pi$. Negative values of the phase are allowed, as the reconstruction can emerge as mirror-symmetrical and complex-conjugated to the original object distribution. In total, 1000 reconstructions were obtained and 50 reconstructions with the least error $E$ were averaged; the resulting reconstruction is shown in Fig. 6(d). The reconstruction of the $512 \times 512$ pixel diffraction pattern should result in bars of width 3, 2.5, 2, 1.5, and 1 pixels, with the thinnest bars of 1 pixel width. This is observed in the reconstruction, where the thinnest bars of 1 pixel width are resolved, see Fig. 6(d).

Next, the simulated $512 \times 512$ pixel diffraction pattern was cropped to $256 \times 256$ pixel in size, as shown in Fig. 6(e) and it was reconstructed in the same manner as before. The resulting reconstruction is shown in Fig. 6(f). The reconstruction of the $256 \times 256$ pixel diffraction pattern should result in the bars of width 1.5, 1.25, 1, 0.75, and 0.5 pixels. Thus, the thinnest and the second thinnest sets of bars should not be resolved. This is observed in the reconstruction, as shown in Fig. 6(f).

# Extrapolation

Each of the 50 reconstructions with the least error $E$ was used for extrapolation. The extrapolation was performed as described above with the object support and the additional restriction on the phase: $-0.2\pi < \varphi < 0.2\pi$. The extrapolated diffraction pattern calculated as squared amplitude of Fourier transform of the average of 50 reconstructions is shown in Fig. 6(g). The average of 50 reconstructions is shown in Fig. 6(h). It is apparent from the reconstruction that all bars up to the second thinnest could be resolved; compare Fig. 6(a) and (h). The phase distribution is also quantitatively correctly reconstructed. Thus, the extrapolation is actually capable of increasing resolution. However, the thinnest bars are still unresolved in the reconstruction of the extrapolated diffraction pattern. Moreover, two bars are observed instead of three bars. This is an artefact that can be explained by the non-perfect initial reconstruction obtained from the $256 \times 256$ pixel diffraction pattern.

## DISCUSSION

## Difference between resolution and detecting fine object features

We would like to note that there is a difference between the resolving power and the recovery of small features in the object from its diffraction pattern. The resolution provided in a diffraction pattern is given by the numerical aperture of the setup; in other words, by the highest detected frequency in the spectrum. On the other hand, a wave diffracted off a small feature is spread over the entire detector area, and therefore its spectrum contains all frequencies, from low (in the central part of the diffraction pattern captured by the detector) to high frequencies (in the outer part of the diffraction pattern not captured by the detector). To demonstrate this observation, we simulated a diffraction pattern of the same object as in the previous example, but with one missing bar, as shown in Fig. 7(a). The missing bar is the thinnest bar in the middle, so its width is at the resolution limit associated with the rim of the diffraction pattern. The difference between the two simulated diffraction patterns, that of the original sample and of the sample without the bar, is shown in Fig. 7(b). This difference demonstrates that in fact, the signal from the thinnest feature is spread over the entire spectrum area. This is a visual example that the possibility of reconstruction of smallest features should not be associated with the resolution provided by the detector area. In addition, this example explains why the extrapolation is indeed possible.

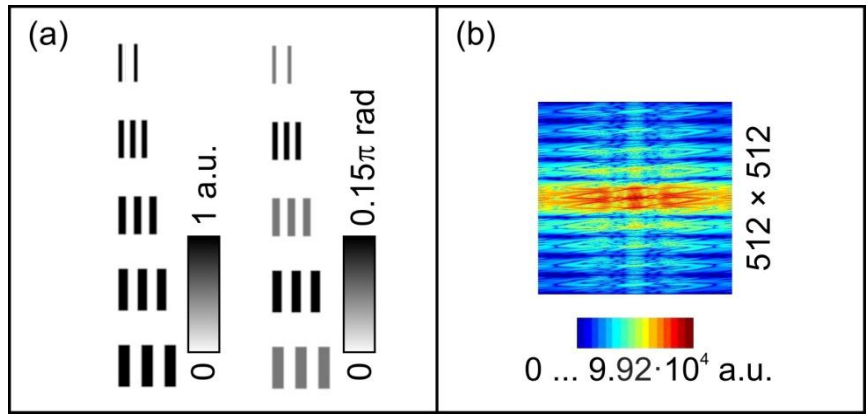

Fig. 7. The effect of fine features in the diffraction pattern. (a) Test object which is the same object as shown in Fig. 6 with only the middle thinnest bar missing. (b) The amplitude of the difference between the diffraction patterns of the original object and the object with one missing bar.

## Zero-padding and extrapolation

We also would like to address the issue of the difference between zero-padding and extrapolation, as the two should not be confused. In zero-padding, an image is extended to an image of a larger size by filling up the added pixels with zeros. When a spectrum of a signal is zero-padded, the signal has a smoother appearance, which can create an illusion of a better resolved signal. In fact, zero-padding of the spectrum does not increase the resolution of the signal; it only achieves a better interpolation of the signal. Zero-padding of a diffraction pattern or a hologram does not increase the numerical aperture of the optical system, and as a consequence the resolution remains the same; a numerical study of that fact has been demonstrated previously (Latychevskaia & Fink, 2013a).

However, in some cases, where the reconstruction procedure requires a single Fourier transform, zero-padding can help to achieve better interpolated reconstructions (Ferraro *et al.*, 2004b, Ferraro *et al.*, 2004a, Ferraro *et al.*, 2007). Also, it has been shown that to solve the problem of abrupt edges of a diffraction pattern, a weighted projection method can be applied to "slightly extrapolate the measured data outside of the computational window"; such a weighting function however "does not allow the extrapolation to extend freely across the entire computational window" (Guizar-Sicairos & Fienup, 2008).

Extrapolation, unlike zero-padding, does not fill up the newly added pixels with zeros; it does however fill them with recovered values that are the continuation of the analytical signal. As a result, the effective numerical aperture of the optical system is increased, and a higher resolution in reconstruction is thus gained.

## CONCLUSIONS

We have demonstrated that a spectrum can unambiguously be extrapolated to a certain level from its fraction, even if only the amplitude of the spectrum is known. We considered the extrapolation applied to the practical problem of coherent diffraction imaging. We showed that the extrapolation ratio, or the amount of signal that can be extrapolated from a portion of the signal, depends on the oversampling ratio, which is the ratio between the total number of pixels and the number of pixels with unknown values in the object domain. The discovered dependency is linear, implying that the higher the oversampling ratio, the greater the range of the spectrum of the object that can be extrapolated.

We showed that an unambiguous extrapolation of the entire complex-valued spectrum from only a section of the spectrum is theoretically possible, although current practical methods are limited to iterative phase retrieval algorithms, which can provide only an approximate solution.

Two numerical examples were studied: a non-binary continuous object and a phase-shifting object. The diffraction patterns of both samples were extrapolated, and the reconstructions of the extrapolated diffraction patterns exhibited an enhanced resolution. However, we observed that, even when the fraction of the spectrum is known with its amplitude and phase distributions, the extrapolated spectrum and the related object are only approximate solutions to the original distributions. Thus, we conclude that the currently available iterative phase retrieval methods are capable of only partially extrapolating the spectrum, and a better extrapolation with a further increase in resolution might be achieved by other superior methods in the future.